\documentclass[manuscript]{acmart} % 

\AtBeginDocument{%
  }

\usepackage{tabularx}
\usepackage{lipsum}
\usepackage{multirow}
\usepackage{color, colortbl}
\definecolor{Gray}{gray}{0.9}
\usepackage{csquotes}

\setcopyright{acmlicensed}
\copyrightyear{2018}
\acmYear{2018}
\acmDOI{XXXXXXX.XXXXXXX}

\acmConference[Conference acronym 'XX]{Make sure to enter the correct
  conference title from your rights confirmation emai}{June 03--05,
  2018}{Woodstock, NY}
\acmISBN{978-1-4503-XXXX-X/18/06}

\begin{document}

\title{Perspectives on Capturing Emotional Expressiveness in Sign Language}

\author{Phoebe Chua}
\email{pchua@nus.edu.sg}
\affiliation{%
  \institution{National University of Singapore}
  \city{Singapore}
  \country{Singapore}
}

\author{Cathy Mengying Fang}
\email{catfang@media.mit.edu}
\affiliation{%
  \institution{MIT Media Lab}
  \state{Massachusetts}
  \country{USA}
}

\author{Yasith Samaradivakara}
\email{yasith@ahlab.org}
\affiliation{%
  \institution{National University of Singapore}
  \city{Singapore}
  \country{Singapore}
}

\author{Pattie Maes}
\email{pattie@media.mit.edu}
\affiliation{%
  \institution{MIT Media Lab}
  \state{Massachusetts}
  \country{USA}
}

\author{Suranga Nanayakkara}
\email{scn@nus.edu.sg}
\affiliation{%
  \institution{National University of Singapore}
  \city{Singapore}
  \country{Singapore}
}

\renewcommand{\shortauthors}{Chua et al.}

\begin{abstract}
Significant advances have been made in our ability to understand and generate emotionally expressive content such as text and speech, yet comparable progress in sign language technologies remain limited. While computational approaches to sign language translation have focused on capturing lexical content, the emotional dimensions of sign language communication remain largely unexplored. Through semi-structured interviews with eight sign language users across Singapore, Sri Lanka and the United States, including both Deaf and Hard of hearing (DHH) and hearing signers, we investigate how emotions are expressed and perceived in sign languages. Our findings highlight the role of both manual and non-manual elements in emotional expression, revealing universal patterns as well as individual and cultural variations in how signers communicate emotions. We identify key challenges in capturing emotional nuance for sign language translation, and propose design considerations for developing more emotionally-aware sign language technologies. This work contributes to both theoretical understanding of emotional expression in sign language and practical development of interfaces to better serve diverse signing communities.
\end{abstract}

% Much progress has been made in our understanding of, and ability to generate, emotionally expressive text and speech. However, research on the expression of emotions in signed languages is limited. To address this gap, we interview 8 sign language users (four Deaf or Hard of hearing (DHH) and four hearing) to understand how they express and perceive emotions in signed languages. We find that....




\received{20 February 2007}
\received[revised]{12 March 2009}
\received[accepted]{5 June 2009}

\maketitle

\section{Introduction}
Sign language users often face barriers to effective communication, particularly in critical settings like healthcare \cite{james2022they} and legal proceedings \cite{taira2019hearing},  where emotional nuance can significantly impact communicated meanings and outcomes. While spoken language users benefit from technologies that are increasingly capable of capturing both verbal content and emotional tone \cite{cowen2019primacy, yang2024emollm}, sign language users still lack access to comparable tools.

Current computational approaches to sign language translation have made progress in capturing both word-level lexical content \cite{li2020transferring} and aspects of continuous signing, such as signing speed \cite{fang2024signllm}. However, extracting precise semantic information from sign language remains challenging due to the richness and temporal complexity of sign language communication \cite{fang2024signllm, liang2024llava, bohavcek2022sign}. While recent work has focused on refining computational approaches to improve translation accuracy, the role of emotional nuance in translation quality remains largely unexplored \cite{qian2023evaluation}.
% Current computational approaches to sign language translation focus primarily on lexical content \cite{fang2024signllm, liang2024llava, bohavcek2022sign}, overlooking the role of emotional nuance in translation quality \cite{qian2023evaluation}. 
This gap can be attributed to several factors, including both an incomplete understanding of emotional expression in sign language and the broader challenge of developing models with nonverbal social intelligence capabilities \cite{li2025mimeqa}. In spoken languages, extensive research has examined how emotion is conveyed not just through words but also \textit{how} something is spoken -- from pitch and intonations to non-verbal expressions \cite{cutler1997prosody}. In contrast, the mechanisms of emotional expression in sign languages remain less well-understood \cite{elliott2013facial,hietanen2004perception,lim2024exploring}.

To develop more effective sign language technologies, it is critical to understand how emotions are communicated in sign language. In the present work, we conducted semi-structured interviews with eight participants, including both Deaf and Hard of hearing (DHH) individuals and hearing signers, from three countries (Singapore, Sri Lanka and the United States). These interviews examined three key research questions: 1) How is emotion communicated in sign language?; 2) What are the cross-cultural similarities and differences in emotional expression across different signing communities?, and 3) How can these insights be translated into design principles for sign language technologies? We analyzed the interview data using thematic analysis methods \cite{braun202418}. Our analysis reveals the richness of emotional expression in sign language through parameters ranging from manual markers like the speed and size of signing, to non-manual elements such as facial expressions and body language. We find that while certain emotion indicators appear universal across cultures, many aspects of sign language are deeply influenced by social and cultural contexts as well as individual differences. The interviews also surfaced several challenges in capturing and translating emotional nuance in sign language interpretation.

In summary, our contributions to theoretical understanding and practical development of sign language technologies are: 1) A cross-cultural examination of emotional expression in sign language, drawing on perspectives from both DHH and hearing signers; 2) Identification of key parameters of emotional expression in signing and how these parameters might be modulated by contextual and individual factors; 3) An outline of specific challenges in capturing emotional nuance for in sign language; 4) Design recommendations for developing more emotionally aware sign language technologies.

\section{Background \& Related Work}
\subsection{Affective prosody in sign language}
The expression of emotion in sign languages has primarily been studied in the domain of sign language linguistics.  Sign languages use visual-spatial mechanisms to express grammatical structure and function \cite{emmorey1993visual}. These mechanisms comprise both manual and non-manual components  \cite{mukushev2020evaluation}.  Manual components involve the hands and are comprised of four parameters: Handshape, Place of Articulation (where the sign is made), Movement (how the articulators move) and Orientation (the hands’ relation towards the Place of Articulation).  Non-manual components include facial expressions, eye gaze, mouthing patterns, and head and body positions. These non-manual components are not merely supplementary, but serve key linguistic and expressive functions and can significantly alter the meaning of signs \cite{tomaszewski2010not}.  For instance, negation of a word (e.g., changing its meaning from HAVE to don't-HAVE)  is often indicated by side-to-side movement of the head \cite{zeshan2004hand}. 

The overlap between linguistic and emotional components in sign language pose significant challenges not just for automatic sign language translation systems, but also for non-signing human audiences. For instance, furrowed eyebrows indicate a content or WH-question in several sign languages but are also typically associated with anger in general human communication \cite{de2009mixed}. Indeed, a recent study found that hearing non-signers frequently misinterpreted linguistic facial expressions as indicators of negative emotions such as sadness or disgust \cite{lim2024exploring}.

While these misinterpretations highlight the complexity of cross-language emotion understanding, they also point to the sophisticated nature of emotional expression in sign languages. Comparable to how spoken languages use vocal intonation, pitch, rhythm or volume to convey emotional content \cite{rao2013emotion, graf2002visual, reilly1992affective}, sign languages employ a rich visual or affective prosody through facial expressions as well as the manual channel, by manipulating parameters such as the tempo, rhythm and size of sign movements. For instance, signs produced in sad emotional conditions exhibit a consistently different quality of movement and duration compared to signs produced in angry conditions \cite{reilly1992affective, hietanen2004perception}. Despite the distinct channels of expression in spoken languages (auditory) versus sign languages (visual-manual), linguistics research suggests fundamental similarities in how they encode emotional information \cite{reilly1992affective}, pointing to potentially universal aspects of human emotional communication regardless of language modality. However, affective computing research centered around sign language and how the rules of emotional displays in Deaf cultures compare to those of hearing cultures is still very limited \cite{orr2024multicollab,gala2017emotional}.

\subsection{Technologies for Sign Language Recognition, Generation, and Translation}
Sign language processing technologies have advanced greatly in recent years, with efforts spanning sign recognition, sign generation (animation), and sign-to-text/speech translation\cite{bragg2019signlanguagerecognitiongeneration}. Computer vision and machine learning have been applied to recognize signs, and sign language translation (SLT) methods typically use either raw image data or skeletal representation of the signer's pose as input. In terms of model architectures, Transformer-based architectures have been used for word-level sign language recognition based on 2D body pose sequence representations \cite{bohavcek2022sign}. The translation capabilities of LLMs also appear to extend to sign languages. SignLLM \cite{fang2024signllm} proposes a framework for transforming sign videos into language-like representations that can easily be passed to off-the-shelf LLMs. In parallel with SLT, research has also investigated sign language production (SLP), which has useful applications such as automatic sign language captioning. SLP typically involves converting text to gloss (a method of sign language transcription), mapping the gloss to pose, then rendering the pose into a video or avatar \cite{fang2024signllm}. Despite significant advances in both SLT and SLP, challenges remain in developing systems that accurately capture and convey emotional nuance in sign language. Most sign language recognition and translation frameworks focus on the literal content of signs and neglect affective signals \cite{lim2024exploringimpactemotionalvoice}.

Several recent approaches have begun to address aspects of this emotional gap in sign language processing technologies. Supplementary emotional voice outputs in sign-to-speech translation were found to improve non-signers' perception of signer emotions, potentially enhancing Deaf to hearing communication \cite{lim2024exploring}. Research has also begun exploring the use of text sentiment analysis of sign video captions to support the generation of emotionally appropriate facial expressions in signing avatars that complement the manual components of sign language production \cite{azevedo2024empowering}. These efforts represent important steps toward more emotionally-aware sign language technologies but remain limited in their understanding of how emotions are expressed across different signing contexts and cultures. We aim to address this knowledge gap through a cross-cultural examination of the key parameters that modulate emotional expression in sign languages. Our work also aligns with the broader call to incorporate sign languages more fully into NLP research \cite{yin2021including}, recognizing them as complete linguistic systems with their own rich mechanisms for conveying both lexical content and emotional nuance.

% Several approaches have been proposed to improve sign language processing. For example, applying tools and theories of Natural Language Processing (NLP) are can help the linguistic modeling of sign languages, and Yin et al. discuss the critical steps needed to include sign language in natural language processing processing\cite{yin2021including}. For emotion recognition, Lim et al. suggested using voice to improve non-signers' perception of signer emotions \cite{lim2024exploring}. Others incorporate sentiment of the witten texts to generating facial expressions to compliment manual components in sign language production \cite{azevedo2024empowering}. Our work contributes to the understanding of challenges in capturing emotional nuance for in sign language and possible strategies to make sign language technologies more emotion-aware. 

\section{Interviews}

\subsection{Recruitment and Participants}
We recruited participants from personal contacts and word-of-mouth in Singapore, Sri Lanka and the US. The inclusion criteria for the study were (i) fluency in at least one sign language, and (ii) being 18 years of age or older. We aimed to include an equal number of native signers and hearing, non-native signers in each country, so as to capture a diverse range of perspectives. 

Eight participants (three female) participated in the study. The study was approved by our Institutional Review Board, and participants were reimbursed \$20 for their time after completing the interview. Participants' experience with sign language ranged from 6-30 years (mean = 16.86). Four participants were native signers who identified as Deaf or Hard of Hearing (DHH), while the remaining participants identified as hearing but had professional experience with sign language (e.g, certified interpreters, teachers of Deaf students). Participants who were Deaf or hard of hearing were asked about their preferred method of communication, and if their preferred mode of communication was sign language, the interview was conducted with the assistance of an interpreter.\\

\begin{table}[h]
\centering
\label{tab:participants}
\begin{tabularx}{\linewidth}{>{\centering\arraybackslash}p{0.03\linewidth}
                              >{\centering\arraybackslash}p{0.08\linewidth}
                              >{\centering\arraybackslash}p{0.13\linewidth}
                              >{\centering\arraybackslash}p{0.15\linewidth} 
                              >{\raggedright\arraybackslash}X}
\hline
ID & Age & Level of Hearing Loss & Preferred Methods of Communication & Sign Language Experience \\ 
\hline
\rowcolor{gray!10} P1 & 30-40 & Profound & Singapore Sign Language (SgSL) & SgSL instructor and interpreter, Deaf arts and music practitioner-artiste; native/fluent signer \\ 
P2 & 30-40 & No loss & English & SgSL interpreter and tutor for DHH students with 9 years of experience \\
\rowcolor{gray!10} P3 & 40-50 & No loss & Sinhala & Principal at a school for DHH students with 25 years of experience \\ 
P4 & 30-40 & No loss & Sinhala & British Sign Language (BSL) teacher at a school for DHH students with 10 years of experience \\
\rowcolor{gray!10} P5 & 30-40 & Profound & BSL & Teacher at a school for DHH students; native/fluent signer \\ 
P6 & 20-25 & Profound & BSL & Student at a school for DHH students; native/fluent signer \\
\rowcolor{gray!10} P7 & 40-50 & No loss & English & American Sign Language (ASL) interpreter in higher education and medical settings with 6 years of experience \\
P8 & 30-40 & Severe & American Sign Language (ASL) & Deaf-of-deaf sign language linguist and lecturer; native/fluent signer \\
\hline
\end{tabularx}

\caption{Interview participants' demographic information.}
\end{table}

\subsection{Procedure}
We conducted semi-structured interviews that lasted between 45-60 minutes via Zoom. During the interviews, participants were asked about the contexts in which they used sign language, and discussed their perspective on the emotional aspects of communication in sign languages. Our questions were grouped into the following three categories:

\begin{itemize}
    \item Perceiving emotional content in sign language (e.g., "When you’re signing with your friends, how can you tell if they are feeling excited, or angry, or sad?")
    \item Expressing emotional content in sign language (e.g., "Do you think that your signing changes when you are experiencing strong emotions?")
    \item Communication technologies for the DHH community (e.g., "What technologies or AI tools do you use to communicate with other Deaf or hearing individuals?")
\end{itemize}

A list of guiding questions we used during the semi-structured interviews can be found in Appendix \ref{appendix}. Additionally, we tailored questions for participants who had a background that might lend an additional perspective to their use of sign language (e.g., as a teacher or performing artist).

\subsection{Analysis}
All interviews were audio-recorded, and preliminary transcripts were generated using Zoom's transcription feature. These transcripts were then manually reviewed, translated into English if necessary, and corrected for accuracy. Following the thematic analysis method outlined in Braun et al. \cite{braun202418}, the transcripts were analyzed in two phases: (i) codebook development through iterative open coding and refinement of emergent themes, and (ii) systematic thematic analysis applying the final codebook to identify patterns and relationships across the interviews.  

In the codebook development phase, three researchers worked independently to code two transcripts. The generated codes were discussed together to identify inconsistencies and come to a consensus about complex nuances in the data. Subsequently, the researchers coded two more interviews independently and met again to discuss and compare the generated codes. 
\section{Findings}
Through the thematic analysis of interviews with DHH individuals, hearing sign language interpreters and teachers of Deaf students,  three major themes emerged around how emotions are expressed in sign language: 1) The fundamental mechanisms through which emotions are conveyed in sign language, 2) how contextual factors and individual differences influence the use of these mechanisms, and 3) the challenges of capturing and conveying emotional nuance during sign language interpretation. 

%four major themes emerged around how emotions are expressed in sign language and the implications for designing accessible technologies. The first three themes focus on the fundamental mechanisms through which emotions are conveyed in sign language, how contextual factors and individual differences influence the use of these mechanisms, and the challenges of capturing and conveying emotional nuance during sign language interpretation. The last theme outlines design recommendations for interfaces to support sign language interpretation and DHH-hearing communication. We present our detailed findings in the following subsections. 

\subsection{Parameters of emotional expression in sign language} \label{sec:parameters}
Regardless of their hearing status or cultural background, participants consistently described emotional expression in sign language as multimodal, combining both manual and non-manual markers to express a rich variety of feelings. First, in terms of manual markers, two parameters that were frequently mentioned were the \textbf{speed and size} of signs, often in relation to changes in emotional arousal. For instance, P1 shared that \textit{"when we're happy or excited, we will tend to sign bigger...we'll [also] move faster, our signs are faster."}  Similarly, P4 observed that if her students were angry, \textit{"their hand movement speed tends to increase, and from that, most of the time, we can identify if they're angry."}

Several participants also brought up how a signer's emotion often influenced their overall \textbf{quality of movement} in a way that was difficult to articulate. P1 provided a comparison between how she signs when she is feeling sad, versus when she is more upbeat -- the signs \textit{"[tend] to be softer...more, like, sloppy."} Echoing this sentiment, P7 said: \textit{"I don't know how else to describe it, other than it just kind of is a register change."} Relatedly, there was a general consensus between participants that learning to express emotions through sign language and read the emotions of other signers was a skill that was learnt intuitively through exposure to the DHH and signing communities. When describing how he learnt to recognize complex expressions such as sarcasm, P7 commented: \textit{"It really just takes being around deaf people...it's something that you do have to get used to and learn. And it kind of becomes more intuitive after a while."}

Third, in terms of non-manual markers, participants emphasized that \textbf{facial expressions} were key to understanding the signer's emotion. When describing how he recognized the emotions of a signer, P5 said, \textit{"facial expressions are the primary way.}" Facial expressions were described as critical for accurately understanding not just a signer's emotions, but also the overall meaning being communicated. P7, a professional interpreter, shared that when interpreting, \textit{"[I'm] picking up signs in the periphery, but really focusing on the face to make sure [I'm] catching the meaning of what is being signed.}" Perhaps due to their linguistic importance, the usage of facial expressions appears to be fairly standardized compared to other aspects of sign language. P8 noted how "there's so many different ways to sign one word, even [just] in ASL, but that facial expression component -- there's not much variation when it comes to that. Many people use them in the same way."

Lastly, aside from facial expressions, \textbf{body language} also emerged as an important source of emotional information, especially to convey more complex or nuanced feelings. P2 illustrated a situation where a signer was upset but trying to control their temper: "You'll be angry, but you also reserve a little bit...it's not just the hands, it's facial expressions; it's even like the way your shoulders move, and things like that." 

%\begin{itemize}
    % \item "For a hearing person who is speaking, you guys have tones, I believe?...But for signing, when we're happy or excited we will tend to sign bigger, and our facial expressions will be more happy. We'll move faster, our signs are faster..." [P1]
    % \item "It's the body. It's not just the hands, it's facial expressions; it's even like, how the way your shoulders [move], and things like that." [P2]
%\end{itemize}

\subsubsection{Emotional expressions are built into the grammatical rules of sign language}
These various parameters of emotional expression are not merely additions on top of sign language -- instead, many participants expressed the view that emotional expressiveness is woven into the grammatical structure of sign language communication. P1 demonstrated two ways of signing the word "sad" in SgSL: first in a standard way, and the second time with greater intensity to convey a meaning closer to "depressed". She explained: \textit{"with my mouth, the way I move, fast or slow, big or small...it's not really so much of cues. It's more that it's part of the sign language."} This sentiment was echoed by P8, who said: "\textit{You can't eliminate the emotions from whatever signs you're producing without omitting some information."} As an illustration, she demonstrated how the sign for "understand" in ASL could be combined with different expressive, non-manual components: "\textit{these different ways that I'm using my facial expressions and shaking my head show you whether I understand, or I don't, or if I understand a little bit. If I do it with no facial expressions, no head shake, no movements, you're not getting the full information.}"

Even though non-manual markers may not be the only way of conveying differences in emotional sentiment or intensity, participants explained that they are often preferred as a more efficient, natural way of communication. For instance, P4 shared: \textit{"When we are teaching, if there’s an adverb [like "very"], we use the specific sign for it and teach it. But in regular conversation, that full sign might not always be used. Instead, [the students'] facial expressions change -- if they’re very happy, their eyes widen; if it’s just happiness, there might be a small smile.}" Similarly, P8 noted that one of the functions of facial expressions was "\textit{becoming an adverb}".

%\begin{itemize}
    % \item "They have the rules in sign language, where we have to express certain things in that way...for example, if I want to give instructions...I'll need to use sharp movements." [P1]
    % \item "There are certain rules...Bigger signs means a bigger word, for example. Eyebrows furrowing up and down, means a question.: [P2]
%\end{itemize}

% \todo{Also mention word choice vs intensity of signs; how emotional expression seems to be more continuous in sign than in spoken languages, formal lexicon sizes tend to be significantly smaller in sign languages compared to spoken languages}

\subsubsection{Emotional expressions specific to sign language}
% \begin{itemize}
%     % \item "Sometimes Deaf people...they will refuse to listen. So what do they do? They'll just close their eyes! [Or] they'll just turn their head away." [P1]
% \end{itemize}
Participants shared ways in which the visual nature of sign language uniquely enabled playful interactions. For instance, P1 gave a humorous example of how a DHH individual might react during an argument: \textit{"[sometimes], they will refuse to listen. So what do they do? They'll just close their eyes and sign...[or] they'll turn their head away."} It also serves as a method of emotion regulation. P6 shared that if he gets angry while in class, he \textit{"asks the other person to stop, and [I'll also] look away."}

Additionally, participants mentioned how native sign language users often demonstrate creativity in their emotional expression through \textbf{ad-hoc modifications of existing signs}. P1 described how \textit{"Deaf people tend to create and add in their own words...for example, when we're excited."} Similarly, P5 noted that \textit{"there are things [the DHH students have] created themselves...they change the sign on their own to show [for instance] anger. Even though there are standard signs, they create their own signs."}

\subsection{Contextual and individual influences on signing styles}
\subsubsection{Conversational context shapes emotion recognition and expression}
Participants frequently discussed how they adapted their signing and emotional expressions based on contextual factors such as the conversational setting, their conversational partner, and the communicative intent. Conversational settings were often described in terms of their \textbf{level of formality}, with participants typically modifying their choice of signs accordingly, For instance, P1 noted that when she was \textit{"just interacting and chit-chatting with someone [versus] a formal, professional sort of setting, the word choices in terms of signs would be very different."} P7 shared a similar perspective, saying:\textit{ "I've interpreted social events...it's a lot more casual [and] laid back...Often in medical settings I go way more formal, like, I use medical terms. I'll kind of gloss over anything graphic."}

The \textbf{fluency} of conversational partners also emerged as a significant factor influencing signing styles. P1 further explained that she tends to emote less when conversing with less proficient signers as it might be confusing for them. To help them understand better, she also mouths her words. 

% Context in terms of:
% \begin{itemize}
%     \item Setting (e.g., formal vs. informal)
%     \item Communication partners (adapting to age groups, fluency levels)
%     \item Communicative intent (arguments, casual conversations, medical appointments)
%\end{itemize}

\subsubsection{Emotional expressiveness beyond everyday communication}
% planning to talk about Deaf arts and music 
When using sign language in artistic or musical contexts, participants discussed how they strategically amplified and stylized their emotional expressions beyond what might be typical in everyday communication. P1 described how she incorporated multiple techniques to convey a rich emotional experience: \textit{"I use different forms of interpreting, like through poetry, visual storytelling, and I’m storytelling also through a visual vernacular technique...I’ll blend it all together to express the song fully, let it out to Deaf people so that they can enjoy the music fully.
}
%For instance, P1 explained that she consciously modulates the intensity of her signs to match the emotional arc of a song: \textit{"When [Beyoncé] was in Dreamgirls, she sings Listen. The storytelling of the girl, you know, who was being bullied by a man? So at first, when you sign 'listen', you have to start small, you sign small, and then you become big..I will develop it that way."} 
Aside from these techniques, P1 emphasized that her approach to musical interpretation places importance on a deep understanding of the content, meaning and structural elements of a song. To develop this understanding, she works together with hearing friends: \textit{"if I'm hearing a song for the first time, in terms of the intonation of the song, I can't really hear. So I'll ask a hearing person first -- how would you interpret [the song] and can you summarize it for me? Once that is done, I'll try to understand the meaning of why the songwriter or the singer is presenting this lyric in this way. [Then] I'll try to interpret and really perform it so that a Deaf audience can enjoy it."}

\subsubsection{Factors that contribute to variations between emotional expressions}
Participants' responses revealed significant variations in emotional expression related to signers' personality, language acquisition background, hearing status, generational norms and cultural context. For instance, in response to a question about how he perceived or understood the emotions of a client when interpreting, P2 answered: \textit{"it's very hard for us to substantiate in concrete terms, because I think a lot of it also depends on the [client's] personality."} He went on to note that some people naturally maintained a more stoic expression. In the same vein, P4 mused: \textit{"different people have different personalities...there's this one student. From the way they behave, it always seems like they're angry. It depends on the individual's behavior."}

Hearing status can also shape emotional expression in sign language. P2 illustrated the hearing bias of hearing sign language interpreters using the scenario of a weather report: "[say] there's a big thunderstorm with CAT 1 lightning...I can either sign SEE, or I might sign BIG STORM NOW...[whereas] a Deaf interpreter...might not even use sign vocab per se, they might go straight to visual gestural communication."

%\todo{TODO: language acquisition background/vocalizations}

Generational differences were also mentioned by several participants as a factor influencing signing styles and choices, with technology playing a notable role in this evolution. P1, who is active on social media platforms such as TikTok, noted a clear difference in the way older generations sign compared to Gen Z youths. P8 offered an explanation for this observation: "\textit{signs have become more efficient so that people can sign faster in smaller spaces. It might be related to technology too -- in iPhones, if we're trying to send a message, we're signing in a smaller space so that people can see all of our message.}" Beyond spatial adaptations for digital platforms, technology is also tranforming how sign language spreads and develops. P8 went on to elaborate: "\textit{traditionally, ASL is passed down through schools and through families...[but now], people are getting that kind of ASL through technology and social media. I think it's impacting the structure...there's a lot of influence from English nowadays.}" Other participants corroborated this technological influence on the transmission of sign language. P6 reported frequently learning new signs through informal interactions with Deaf friends rather than in a classroom setting, while P3, a school principal, shared: \textit{"When I use an old sign that I'm familiar with, the students ask me what it is. And when I ask if they don’t know it, they teach me the new sign they’re using in place of the old one I use."}

Finally, cultural context also influenced emotional expression in sign language. P1 explained: "In Singapore, your face tends to go upwards [when using Singlish phrases like \textit{wah lau}]. So for Deaf people, we tend to follow [the head movements and body language] as well." P7 added that in his experience, sign language users "definitely do take body language from English speakers to apply to signs, and sometimes it keeps the same meaning, and sometimes it changes slightly", illustrating how body language and communication norms flow and adapt across hearing and Deaf communication cultures. 

% Variations in terms of:
% \begin{itemize}
%     \item Sign language acquisition background (e.g., native sign language users, oral upbringing)
%     \item Generational differences in signing styles and word choices 
% \end{itemize}

\subsection{Challenges in sign language interpretation and translation}
\begin{displayquote}
As an interpreter, we are purely just the mouth of a Deaf person, and the hands of the hearing. --P2
\end{displayquote}

A key challenge in language translation is conceptual equivalence -- providing a translation that is technically and conceptually accurate, beyond a simple literal translation \cite{squires2009methodological}. Our interviews revealed several ways in which this challenge is amplified when trying to capture and communicate emotional nuances during sign language interpretation. 

\subsubsection{Layers of translation loss}
Several interpreters described how meaning and emotional nuance can be lost due to the bias of hearing or non-native sign language interpreters. P2 highlighted how "\textit{for hearing people, when we sign, we have a hearing bias. So sometimes we would phrase it in a way that makes more sense for a hearing person to understand grammatically...it's not as visual as it could have been.}" The challenge of translating sign language in a conceptually accurate manner becomes particularly apparent with emotionally complex content. In the case of humor, P7 noted: "\textit{it's such a challenge to take a joke in English and interpret it into ASL and have it come across the same -- you have to do a lot of tweaking.}" P7 also shared the challenge of translating in emotionally charged situations, which require a delicate balance of conceptual accuracy, empathy and dignity: "\textit{one emotion that I have a difficult time with, and I think most interpreters would agree...if a patient is at the hospital, if some is crying and very upset, I'm not going to voice it with a sob in there, because [the medical staff] can see that the patient is crying. So I'll maybe kind of lower my voice and sound a little bit more sad and frustrated.}"

To overcome these challenges, interpreters used several strategies. First, \textbf{gathering contextual information} helped them to calibrate their tone and expression to match the communicative intent of their clients. P2 explained, "\textit{As an interpreter, you kind of need to know the more bird's eye view of what's happening, and change the expression and tone based on what the context is.}" He went on to provide an example: "\textit{If a subordinate wants to confront his boss, the tone can't be neutral, right?}" Relay interpretation, where a Deaf interpreter serves as an intermediary between the hearing interpreter and the Deaf audience, adds another layer of translation but potentially helps to communicate meaning and emotional nuance in a more authentic manner, as P2 also explained: "\textit{A Deaf interpreter would be a second bridge...they take away the hearing bias or the hearing tone.}" Deaf participants also mentioned self-regulation strategies they had developed to communicate with hearing interpreters more effectively. For instance, an observation that frequently emerged during our interviews was how Deaf individuals tended to sign rapidly during arguments. P1 reflected on this tendency, saying: "\textit{There will be a bit of a problem, because the interpreter might not be able to catch everything. Right now, I'm still learning that sometimes, if you're feeling angry, we need to control."}

\subsubsection{Interpreter constraints}
Physical and cognitive fatigue emerged as a significant constraint on interpreters, especially when interpreting for an extended period of time (e.g., beyond 45 minutes). P7 shared that in these situations, a team typically works together: "\textit{There's a lot of mental fatigue in interpreting. After about 20 minutes or so...your articulation and accuracy decreases, so we switch on and off.}"

\subsubsection{Additional challenges introduced by technological mediation} \label{sec:challenges}
The rise of smartphones, video remote interpreting (VRI) services and on-screen sign language interpretation has helped to increase access to audiovisual media \cite{bosch2020sign} as well as important services such as healthcare. However, a recent study found that less than half of users were satisfied with the quality of VRI services \cite{kushalnagar2019video}. Participants in our study echoed these concerns, highlighting several specific challenges they encountered with video-mediated sign language communication. 

First, \textbf{spatial cues and subtle information tends to be lost over video.} For instance, P7 noted: "\textit{ASL is visual and three-dimensional. So, some of the non-manual stuff gets lost on video, like some of those subtle facial expressions [and body movements].}" He then demonstrated how parts of the hand could be occluded when one is signing while directly facing the camera, explaining: "\textit{sometimes you have to rearticulate the sign so you can see it more clearly. [The video appears] so flat that you can miss a handshape.}"

Secondly, participants commented that \textbf{video size, quality and signing speed} greatly impacted the usability of video-mediated accessibility services and communication. For everyday communications, P6 indicated a preference for face-to-face interactions: "\textit{It is easier for me. However, since this video call screen is large, I'm okay with it as long as I can clearly see you.}" News interpretation presented a more serious challenge, with P6 explaining: "\textit{The signs are not visible enough, and the speed of the signs is also an issue. It's not a fault with the interpreter -- the person is trying to match the speed of the news reader.}" P4 agreed with this assessment, saying: \textit{"I can't properly see the signs in the news. Here [in Sri Lanka, the news only allocates] a very small space for the interpreter...the signs aren't clear enough."}  The combination of small screen space, lack of high-fidelity video and rapid signing speed required to keep pace with spoken news delivery appears to create substantial barriers to comprehension even for native and fluent sign language users, and especially so for members of the Deaf community who may have visual impairments.

    % \item \todo{Include quote from Khanishka's interview where he mentioned not being able to understand sign language captions on the news}
    % \item Screen limitations (e.g., interpreter given too little space on screen)
    % \item Loss of 3D informtation over a video screen

%\section{Design Considerations for sign language interpretation technologies} \label{sec: recommendations}
\section{Considerations and opportunities for designing emotionally-aware sign language technologies} \label{sec: recommendations}

% Designing Emotionally-Aware Sign Language Technologies: Challenges, Considerations and Opportunities
%  There is a difference between signing as information and signing as communication
Participants' responses highlighted several key challenges in sign language interpretation: the need to account for subtle cues in facial expressions and body language, the influence of context and individual differences on signing styles, and the difficulties encountered by DHH individuals and interpreters when using and providing accessibility services. Our findings revealed important design considerations, including feedback mechanisms, contextual appropriateness, visual accessibility, and the need for co-development with the Deaf community. 
%In this section, we discuss the implications of these findings for the development of sign language technologies, integrating additional interview excerpts where they illuminate design opportunities. 

% this section looks and feels very wordy -- wonder if it's possible to break it up with a diagram somehow? 

%These considerations fall into four main categories: 1) Incorporating feedback mechanisms (i.e., Deaf person can verify that the translation is accurate); 2) Contextual appropriateness (first: human translation is preferred in important/sensitive scenarios, like medical diagnosis, less important for repetitive work. second: emotional accuracy can be important, but sometimes translation accuracy is critical, like in medical scenarios/when giving a lecture and you want to be precise about terminology); 3) Visual accessibility considerations (high-quality video on a large-enough screen), 4) Thinking out of the box re. sign language translation: for example, things like visual aids may be much more helpful for an interpreter than a technology that directly does the translation for them.

\subsection{Incorporating feedback mechanisms that allow users to verify translation accuracy}. 

\begin{displayquote}
I think it's good to have, you know, more of a collaborative approach, where having an interpreter is good. So that [as] a Deaf person, you can really be involved in the hearing world and the hearing person also can be better involved in the Deaf world. --P1
\end{displayquote}

A key advantage of using human interpreters instead of digital interpretation tools is their ability to provide and receive immediate feedback. This feedback operates on multiple levels. First, it \textbf{allows signers to verify if their intended meaning and emotional tone are being conveyed accurately}, by observing the interpreter's facial expressions, body language and lip movements. As P2 explained in detail: "\textit{I think the reason why a lot of Deaf people still prefer [human] interpreters is because there is a feedback loop...if you are [in a confrontation], and you look at your interpreter and they're like, smiling, you kind of know that they're not matching what you're trying to say correctly. Whereas for technology, you might not know whether number one, the content is correct and number two, if the emotion and energy is correct.}" Additionally, it also \textbf{enables interpreters to check if they have properly understood a signer} and seek clarification when needed. P2 also shared his experience as a tutor, saying: "\textit{One word can have a lot of different signs. So you need to...process it and ask [the student], is this what you actually mean?}" 
The collaborative, bidirectional nature of the feedback process helps provide reassurance that communication aligns with the signer's intentions. However, current work on sign language translation interfaces has yet to consider incorporating the ability for users to provide or receive feedback \cite{prietch2022systematic}. By including explicit feedback channels, designers could help users better understand how their signing is being interpreted, both in terms of content and emotional register. This could be done through methods such as real-time visualizations of detected emotional states \cite{liu2021significant}, providing confidence scores for sign recognition \cite{sun2024adaptive}, or interactive tools that enable users to correct or clarify interpretations.

\subsection{Contextual appropriateness of sign language technologies}
% 2) Contextual appropriateness (first: human translation is preferred in important/sensitive scenarios, like medical diagnosis, less important for repetitive work. second: emotional accuracy can be important, but sometimes translation accuracy is critical, like in medical scenarios/when giving a lecture and you want to be precise about terminology)
% \begin{itemize}
%     \item "If a hearing person is trying to learn [sign language with] AI tools, I'll go against it. I'll try learning the signs by communicating slowly, face to face." [P1]
% \end{itemize}

\subsubsection{Technologies for automatic sign language translation and interpretation.}

In terms of automatic sign language translation technologies, participants discussed how their appropriateness varies by context, particularly in terms of the \textbf{stakes and complexity of the communication involved.} P1 noted that such tools might be suitable for routine, repetitive communication: "\textit{[if] the interpreter has to keep coming down to do the same thing over and over again, communicating the same thing...then, you know, what's it for?} " P7 expressed a similar sentiment: "\textit{there's a patient who goes [to the hospital] almost every day for chemotherapy infusions, and it's the same thing every day. She just sits there for an hour, an hour and a half, so she doesn't care if the interpreter's in person or not.}" This suggests taking a strategic approach when deploying sign language translation tools -- utilizing them for routine communications, so that human interpreters (who are often in high demand) can be allocated to more complex, high-stakes situations, such as "\textit{[medical] appointments where there's information, and possible diagnoses}" [P7]. As P1 added, "\textit{because we don't really have a lot of sign language interpreters, we have to make sure that the level of work matches to the interpreter.}" The \textbf{tolerance for translation errors in terms of word choice and emotional tone also varies by context and user intent}. P7 commented that he had rarely been corrected by clients for his word choice in medical settings, saying: "\textit{if they are going to a doctor, they're probably going to care a bit less about word choice and more about what their problem is.}" However, he added that there are contexts where precise terminology is more important, such as "\textit{a Deaf lecturer...they [might] want to be very specific about what they're saying, so they might correct the interpreter in that situation.}" Similarly, P2 described situations in which accurately conveying emotional tone is key to "\textit{the accuracy of the message}", such as when interpreting arguments or apologies. 

These insights suggest that designers need to carefully consider use-case boundaries when deploying automatic sign language translation capabilities, keeping in mind both their current and future levels of accuracy. Translation systems might be suitable in low-stakes, routine situations where communication patterns tend to be more predictable. However, for more complex communications such as medical diagnoses, legal proceedings or academic settings, the limitations of translation systems should be explicitly acknowledged, and human interpreters should remain the primary choice. Building on the earlier design implication of including explicit feedback channels that enable users to correct or clarify interpretations, future translation systems could adopt context-aware features that adapt to different settings. For instance, as communication complexity increases, systems might offer more detailed translations, present alternative interpretations for users to select from, or allow for real-time refinement of word choice. 

\subsubsection{Technologies for sign language learning.}
Aside from supporting communication, participants also discussed the use of technology-based tools for sign language learning, such as mobile apps. Despie their growing popularity \cite{david2024sign}, most participants expressed skepticism about their effectiveness as a primary learning method. P1 was particularly direct: "\textit{If a hearing person is trying to learn [sign language with] AI tools, I'm against it. I'll try learning the signs by communicating slowly, face-to-face.}" This preference for in-person learning appears to stem partially from the dynamic nature of sign languages, which, like spoken languages, are growing and evolving over time \cite{bragg2019sign}.  P2 noted that even with more than a decade of interpretation experience, he is "\textit{still learning new things, like Singaporean slangs, for example, and it's constantly changing}." He went on to emphasize the importance of learning sign language from the Deaf community:  "\textit{The best way to learn is kind of just jumping in with the Deaf of asking, 'what are you signing?}'" P4, who also has a decade of experience teaching at a school for DHH students, similarly noted that when new signs emerge, her DHH colleagues and students "\textit{are aware of it right away. But for us [hearing teachers], it takes a bit more time to catch up.}" Learning from the Deaf community is particularly important for communicating emotions in sign language, which many participants reported to be a skill they picked up largely intuitively. In talking about the learning process , P7 commented: "\textit{I mean, it's hard. It's not something you learn in classroom at all, it's really impossible to learn stuff like that. It really just takes being around Deaf people.}" Overall, our interviews suggest that technology-enabled sign language learning tools should be viewed as a complement to, rather than a replacement for, the immersive social aspects of sign language learning that appear crucial for developing true conversational fluency.

\subsection{Visual accessibility considerations}
Visual clarity emerged as a crucial consideration for sign language technologies. Based on the challenges discussed in Section \ref{sec:challenges}, we propose three key design considerations for accessibility services such as video remote interpreting (VRI) and on-screen sign language interpretation. First, such services should aim to provide high video resolution and ensure that adequate screen space is allocated to the signer, so that viewers are able to clearly see both manual gestures and non-manual markers such as facial expressions and head movements. Second, future work could explore methods to preserve or highlight subtle facial expressions and movements that tend to be easily lost in video translation, such as video face filters that place visual emphasis on important non-manual markers \cite{herring2024strategic}. Finally, researchers could investigate ways to capture and present the three-dimensional nature of signing through solutions such as multiple viewing angles, or three-dimensional avatars.

%\subsection{Beyond sign language recognition, generation and translation}
\subsection{Opportunities to advance technologies for sign language recognition, generation and translation}
% 4) Thinking out of the box re. sign language translation: for example, things like visual aids may be much more helpful for an interpreter than a technology that directly does the translation for them.

Research on sign language technologies predominantly focuses on sign language recognition, generation and translation \cite{bragg2019sign}. However, participants highlighted practical issues and opportunities that suggest value in alternative technological approaches to support sign language interpretation. First, visual resources often prove more valuable than direct translation tools. P7 noted, "\textit{ASL is so visual, so any visual aids are super helpful."} This is especially relevant in contexts like STEM education, where complex or abstract concepts need to be communicated while many terms still lack standardized signs \cite{yin2024asl}. As P2 explained, "so what if I spell 'endocrine system'? [It's still difficult to understand] what the meaning is, right?" These insights suggest that providing visual context and reference materials alongside sign language interpretation may help signers convey meaning more effectively, than if efforts were to focus solely on translation capabilities.

Second, our findings suggest avenues for improving computational methods to capture emotional nuance in sign language. Emotions are often conveyed through modifications to signing parameters rather than distinct lexical items -- in the words of P8, "\textit{You can't separate these emotions or facial expressions from the signs themselves, the actual manual production of the words.}" 
In addition, as discussed in Section \ref{sec:parameters}, emotions are conveyed through variations in the signing speed and quality of the signs, similar to what Reilly et al. found \ref{reilly1992affective}.
Future systems might therefore benefit from methods capable of capturing spatial and temporal information, such as optical flow analysis \cite{ding2024language}, to account for how markers such as signing speed and size convey emotional intensity. Additionally, given participants' emphasis on the critical role of non-manual markers, researchers should also consider more sophisticated approaches to extracting and representing facial expressions \cite{lian2025affectgpt} and body pose \cite{yin2025smplest}. Lastly, these technical capabilities should be developed in tandem with context awareness so that emotional parameters can be interpreted and generated in a suitable manner, as our findings revealed notable contextual variations in how and when emotionally expressive aspects of sign language are used.

\subsection{Involving the DHH community in technology development}
Our findings strongly indicate that sign language technologies must be developed with the Deaf community as co-creators \cite{spinuzzi2005methodology}, not merely end users. Participants consistently rejected the conventional paradigm where accessibility solutions position DHH individuals as passive recipients requiring assistance. P8 articulated this point directly, saying: "\textit{They've put a burden on the Deaf community in the end, for them to facilitate communication with hearing people.}" This burden-shifting could explain why many well-intentioned technologies face resistance in the wild. As P2 noted: "\textit{Sometimes when it's packaged in a way like, 'We're here to help Deaf people', it's like -- do I really need help? Especially if the technology is too basic.}" These insights suggest the value of a fundamental change in orientation when developing sign language technologies: rather than only creating tools that help DHH invididuals adapt to hearing norms, designers should also look towards creating technologies that enable hearing users to understand sign language. This inverted accessibility model supports bidirectional communication where both parties are actively sharing the responsibility to create successful interactions. P4 envisioned the impact of such a change:  "\textit{Usually, when [one of our Deaf students] answers a question from a hearing person, they don't initiate further conversation. But if there's a tool that helps the hearing person understand what the Deaf person is signing, there would be fewer issues...both should have some kind of connection for effective communication.}"

\section{Limitations \& Future Work}
In the current study, we conducted exploratory qualitative research to identify critical technical and design considerations for developing emotion-aware sign language technologies. Though our findings shed light on the key parameters of emotional communication in sign language, there are some limitations to this work. First, we recruited participants across three countries to capture diverse perspectives and surface cross-cultural differences in emotional expression or perception. However, the small sample size could limit how well the findings represent any single cultural context. Although the semi-structured interviews averaged 45-60 minutes, the emphasis on emotional expressiveness meant our findings may have overlooked other important aspects of sign language communication that could impact technology design. 

We observed that emotional expression appears more continuous in sign language compared to spoken language. For instance, while English speakers might use distinct terms like "angry" versus "furious" to communicate varying intensities of displeasure, signers may modify a single base sign through parameters such as speed or size to convey similar gradations of emotion. This pattern could be partially explained by the generally smaller formal lexicon sizes of sign languages \cite{caselli2017asl}. However, it also has interesting implications for research on emotion regulation and processing, particularly given evidence linking emotional vocabulary to emotion regulation capabilities \cite{tugade2004psychological, barrett2017theory}. Future work could examine whether this more continuous approach to emotional expression in sign language influences how native sign language users conceptualize and regulate their emotions, compared to native spoken language users. 

% We also did not isolate the unique attributions of emotions that are present within sign language. It is hard to do it because we cannot explicitly ask if a signer to compare their emotional expressions between when they are signing vs when they are speaking. More specifically, we did not have a quantified way of comparing the emotional displays between sign languages and other spoken languages because we rely on first-person experiences.

Finally, many participants had rich experiences in specific signing contexts beyond everyday communication (e.g., Deaf arts and music); more in-depth interviews about these specific use-cases could help to further inform the design of more specialized sign language technologies. Additionally, the current qualitative study focused on gathering perspectives rather than testing specific technical approaches, limiting our ability to gather concrete feedback on a proposed solution. Future work should consider developing prototypes and conducting user studies to test the design recommendations outlined in Section \ref{sec: recommendations}.

% \begin{itemize}
%     \item Acknowledge concerns regarding ethical and responsible development of sign language technologies \cite{de2021good}
%     \item Interviews align with findings from the literature:
%     \begin{itemize}
%         \item Common misconception that sign languages are primarily articulated by the hands is incorrect, and signers primarily focus their attention on each others' face \cite{pfau2010nonmanuals}
%         \item Observed that emotional expression appears to be more continuous in sign language versus spoken language (e.g., English may have distinct words like "angry" vs "furious" that express the same underlying emotion with differing intensity, while in sign language the base sign might be the same, just a change in sign intensity). One possible reason for this observation is that formal lexicon sizes tend to be significantly smaller in sign languages compared to spoken languages \cite{caselli2017asl}
%         \begin{itemize}
%             \item has interesting implications for research given emerging evidence that emotional vocabulary influences emotion regulation ability
%         \end{itemize}
%     \end{itemize}
% \end{itemize}
\section{Conclusion}
We conducted semi-structured interviews with DHH and hearing signers across three countries to explore how emotions are communicated in sign language, and the cross-cultural similarities and differences in emotional expressions. Our findings revealed that while manual parameters carry emotional information, non-manual parameters such as facial expressions, head movements and body language play a critical role in conveying both the lexical and semantic intent of the signer. We provide insights on the opportunities and challenges around capturing emotional nuance in sign language and make recommendations for future research and design of sign language technologies, including considering use-case boundaries and incorporating feedback loops for automatic sign language translation systems, and encouraging a participatory design approach.

\begin{acks}
\end{acks}

\bibliographystyle{ACM-Reference-Format}
\bibliography{references}

%%%%%%%%%%%%%%%%%%%%%%%%%%%%%%%%%%%%%%%%%%%%%%%%%%%%%%%%%%%%

\newpage
\appendix

\section{Semi-structured interview questions} \label{appendix}
\textbf{General:}
\begin{itemize}
    \item  Can you describe the different contexts in which you typically use sign language (e.g., everyday personal communications, professional settings)? \\
\end{itemize} 

\textbf{Perceiving emotional content in sign language:}
\begin{itemize}
    \item How do you interpret/understand what emotional nuances are being signed by others?
    \item What markers or cues (e.g., facial expressions, body language) are important for conveying emotion?
    \item Can you give examples of how the same sign can change in meaning based on emotional context?
    \item Have you ever misunderstood the sentiment someone was trying to express? If yes, could you provide an example of the scenario? \\
\end{itemize}

\textbf{Expressing emotional content in sign language:}
\begin{itemize}
    \item Do you think that your signing changes when you are experiencing strong emotions (e.g., anger, sadness)? If so, how?
    \item When you are consciously trying to express emotions in sign language, how do you do so? Are there any emotions that are particularly challenging to express? \\
\end{itemize}

\textbf{Communication technologies for the DHH community:}
\begin{itemize}
    \item How do you think digital technologies are shaping your experiences as a sign language user?
    \item Have you used any technologies or AI tools to communicate with DHH/hearing individuals? If so, can you give us an example of these technologies? 
\end{itemize}

\end{document}